Notice: This manuscript has been authored by UT-Battelle, LLC, under Contract No. DE-AC05OOOR22725 with the U.S. Department of Energy. The United States Government retains and the publisher, by accepting the article for publication, acknowledges that the United States Government retains a non-exclusive, paid-up, irrevocable, world-wide license to publish or reproduce the published form of this manuscript, or allow others to do so, for the United States Government purposes. The Department of Energy will provide public access to these results of federally sponsored research in accordance with the DOE Public Access Plan (http://energy.gov/downloads/doe-public-access-plan).

# Probing temperature-induced phase transitions at individual ferroelectric domain walls

Kyle P. Kelley,[1,a] Sergei V. Kalinin,[1] Eugene Eliseev,[2] Shivaranjan Raghuraman,[1] Stephen Jesse,[1] Peter Maksymovych,[1] and Anna N. Morozovska[3,b]

[1] Center for Nanophase Materials Sciences, Oak Ridge National Laboratory, Oak Ridge, TN 37831
[2] Institute for Problems of Materials Science, National Academy of Science of Ukraine, Krjijanovskogo 3, 03142 Kyiv, Ukraine.
[3] Institute of Physics, National Academy of Science of Ukraine, pr. Nauki 46, 03028 Kyiv, Ukraine.

**Abstract:**
Ferroelectric domain walls have emerged as one of the most fascinating objects in condensed matter physics due to the broad variability of functional behaviors they exhibit. However, the vast majority of domain walls studies have been focused on bias-induced dynamics and transport behaviors. Here, we introduce the scanning probe microscopy approach based on piezoresponse force microscopy (PFM) with a dynamically heated probe, combining local heating and local biasing of the material. This approach is used to explore the thermal polarization dynamics in soft $Sn_2P_2S_6$ ferroelectrics, and allows for the exploration of phase transitions at individual domain walls. The strong and weak modulation regimes for the thermal PFM are introduced. The future potential applications of heated probe approach for functional SPM measurements including piezoelectric, elastic, microwave, and transport measurements are discussed.

[a] kelleykp@ornl.gov
[b] anna.n.morozovska@gmail.com

Ferroelectric domain walls separate regions of uniform polarization in ferroelectric materials and have recently been investigated for the broad range of functional behaviors they may exhibit.[1] Following the discovery of the domain wall conductance by Seidel,[2] the electronic properties of the ferroelectric domain walls have been explored by multiple groups.[3-8] These studies targeted both the fundamental mechanisms of domain wall conduction,[9] as well as applications in reconfigurable oxide electronics.[5] The critical enabling element was local scanning probe microscopy techniques that allowed probing the properties of individual domain walls,[10-12] as well controllable manipulation of ferroelectric polarization and the creation of specific ferroelectric[13, 14] and ferroelastic[15] domain structures.

Similarly, advancements in (Scanning) Transmission Electron Microscopy have enabled direct observations of the atomic structures in the vicinity of domain walls and topological defects. In this, direct visualization of atomic column positions and shapes allows the reconstruction of symmetry breaking distortions on a unit cell level.[16-21] Here, the direct evidence towards the emergence of phases associated with novel order parameters,[22] polarization rotation,[23] and carrier accumulation[24] have been observed.

From the theoretical perspective, the emergent properties of the domain walls can be generally traced to the symmetry breaking phenomena emerging when the primary order parameter becomes zero.[3, 25] Under these conditions, the phases and structural behaviors associated with secondary order parameters including magnetic,[26] ferroelastic,[27] or superconductive[28] can emerge. In the systems close to the phase coexistence, the walls can precipitate transition to the spatially-modulated phases induced by flexoelectric interactions.[29, 30] For the walls associated with the discontinuity of polarization vector, the additional mechanisms of emergence of new functionalities is the coupling with the semiconductor system of the material.[31] Finally, strain effects at the walls can lead to oxygen vacancy or other mobile charge accumulation and associated structural and electronic responses.[32-34]

However, until now, the vast majority of domain wall studies via scanning probe and electron microscopy have been limited to the constant temperature well below the ferroelectric transition. For the few cases when variable-temperature behaviors of surface potential or piezoresponse force microscopy (PFM) signals were explored,[35-40] these studies were performed under uniform heating conditions, by changing the temperature of the whole sample. Little is presently known about local temperature-induced phenomena at the domain walls, such as possible local temperature induced phase transition and material responses under the conditions of strong thermal gradients.

Here we introduce a new approach, UV laser assisted thermal band excitation PFM (tPFM), to probe thermal- and voltage dynamics in functional materials based on the synergy of local heating and local biasing. By aligning a UV laser normal to an atomic force microscope cantilever while employing band excitation piezoresponse spectroscopy (Figure 1a), the local ferroelectric and mechanical properties as a function of temperature can be measured simultaneously, providing insight into the local nature of the transition. Specifically, we apply a triangular voltage wave (here chosen as 0-4.75 V, 1 Hz) to the UV laser while measuring the PFM amplitude, phase, and resonant frequency in a point-by-point spectroscopy framework. In this manner, functional responses of the material are probed as a function of local heating where the relatively small tip-surface junction (radius less than 25 nm) mitigates long range heating. To complement local studies, uniform heating is employed to evaluate thermal properties across the transition without the presence of a large local thermal field created in tPFM.

As a model system, we have chosen uniaxial $Sn_2P_2S_6$ (SPS) bulk single crystals which undergo a ferroelectric-paraelectric phase transition at approximately 338 K.[41] The low Curie temperature makes this system ideal for temperature dependent piezoresponse force microscopy measurements. Here, we aim to explore the local nano- and micro-scale ferroelectric properties and domain wall dynamics of single crystal SPS as a function of temperature via tPFM and uniform heating techniques. Combined imaging and spectroscopic studies have enabled us to determine the nature of the propagating phase boundary with increasing temperature, the heterogeneous nature of the transition, and how thermal gradients affect the functional properties near the phase transition.

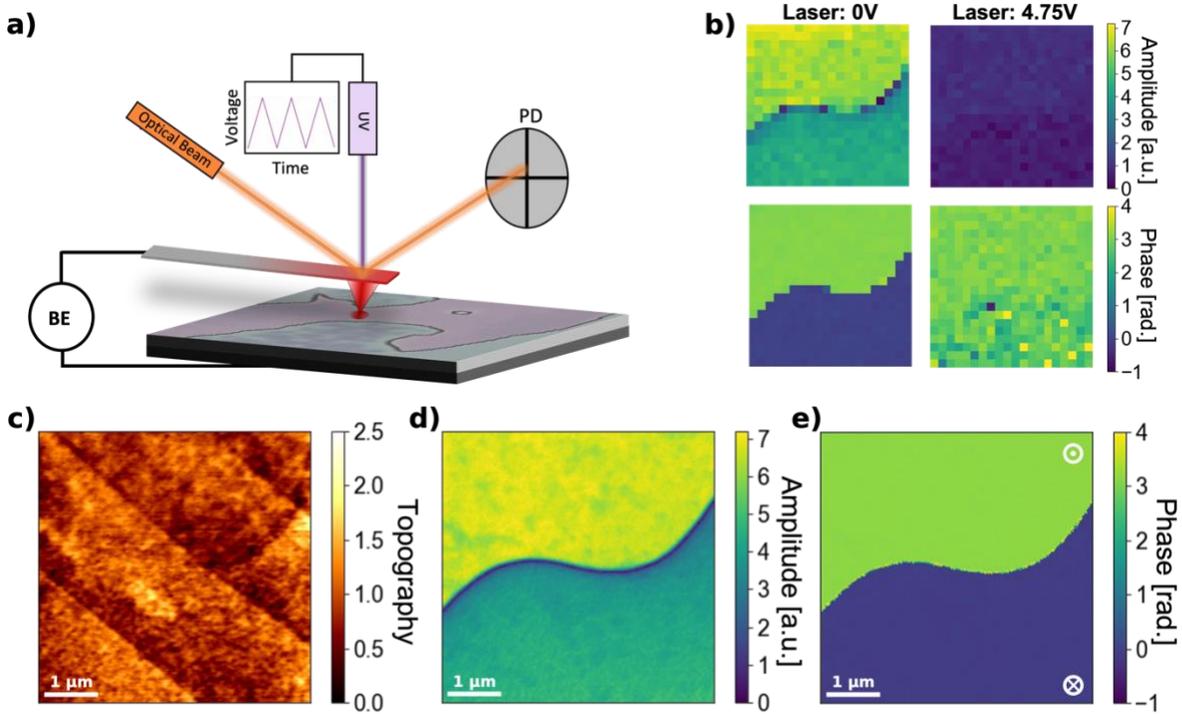

**Figure 1: Experimental setup of thermally assisted band excitation piezoresponse force microscopy.** a) Experimental illustration with ultra-violet laser aligned to AFM probe normal for thermal cycling. Triangle wave is applied to laser input for cantilever heating while ferroelectric and mechanical properties are measured via band excitation PFM (optical beam). b) Band excitation piezoresponse spectroscopy amplitude (top panels) and phase (bottom panels) spatial maps selected at a laser voltage of 0V (left, room temperature) and 4.75V (right) over a 20x20 grid. $Sn_2P_2S_6$ bulk single crystal c) surface topography, band excitation piezoresponse force microscopy d) amplitude and e) phase, where dark blue and light green phase regions correspond to down and up polarized domains, respectively.

As shown in Figure 1c, SPS possesses a smooth step and terrace-like surface morphology ideal for SPM-based techniques. Band excitation PFM amplitude and phase images at room temperature of 180° domains are shown in Figure 1d,e respectively. As such, a clear domain wall can be observed separating the two 180° domains, which is central to this study. We then apply thermal spectroscopy as shown in Figure 1b rastering from the top left and ending in the bottom right corner. At a laser voltage of 0V, i.e. room temperature, the spectroscopy amplitude and phase (left panels, top and bottom respectively) clearly resemble the PFM images (Figure 1d,e). However, at the largest laser voltage, i.e. 4.75V, a strong reduction in response and relatively uniform phase image is observed (Figure 1b, right panels), indicating probe temperatures above the Curie temperature and a corresponding reduction of piezoelectric signal due to the local phase transition.

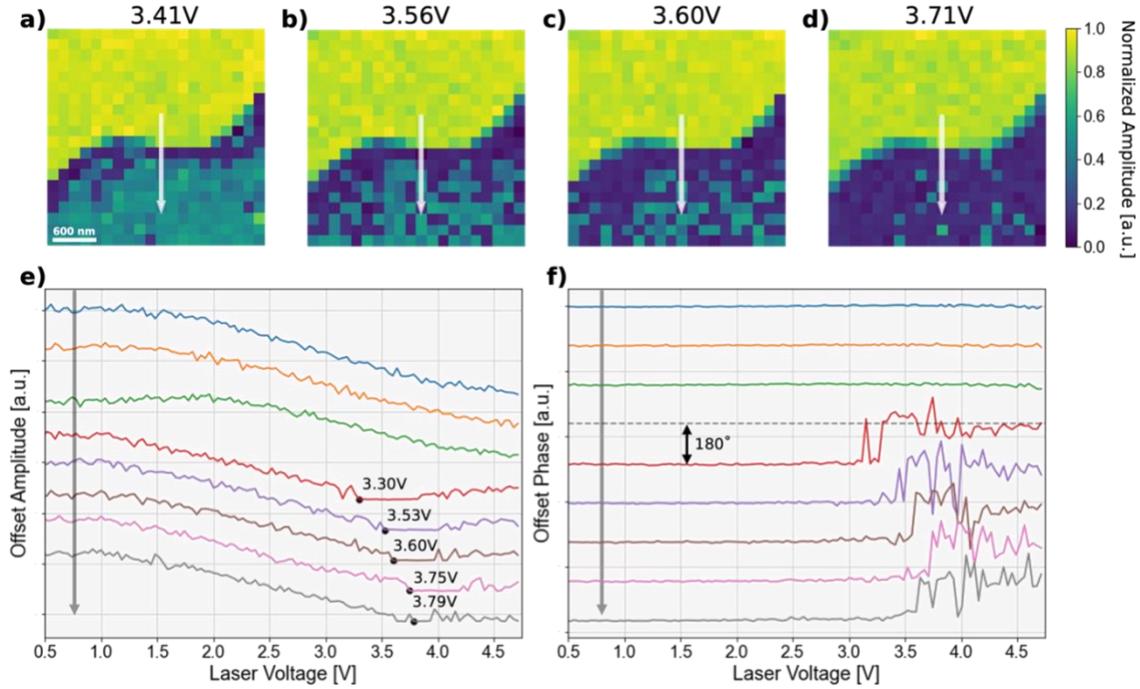

**Figure 2: UV laser assisted thermal band excitation spectroscopy.** Spatial maps of normalized tPFM amplitude extracted at a laser voltage of a) 3.41 V, b) 3.56V, c) 3.60V, and d)3.71V. Note, panels (a-d) are slices of tPFM 3D dataset at specific laser heating voltages. A clear phase transition is observed propagating away from the domain wall versus laser voltage, i.e. temperature. e) Waterfall plot of tPFM amplitude and corresponding f) phase acquired across domain wall (white arrow panel a-d) illustrating the laser voltage at which the amplitude drops to zero increases further from the domain wall.

To access the domain wall behavior near the phase transition temperature, we plot the normalized tPFM amplitude per laser voltage near the transition onset (Figure 2a-d), which occurs between 3.41V and 3.71V. Initially, the domain wall begins to broaden (Figure 2a) with increasing temperature, while further increasing the temperature results in a monotonic decrease in response only for the lower domain (Figure 2b-d) located at the bottom of the image corresponding to polarization pointing into the plane. Interestingly, we observe distinctly different phase transition behaviors between the domains. To examine this further, we plot spectra acquired across the domain wall as a function of laser voltage (Figure 2e,f), i.e. increasing temperature, acquired from the white line in Figure 2a-e. For clarification, the top and bottom curves in Figure 2e,f correspond to spectroscopic locations indicated by the beginning and end of the white arrows presented in

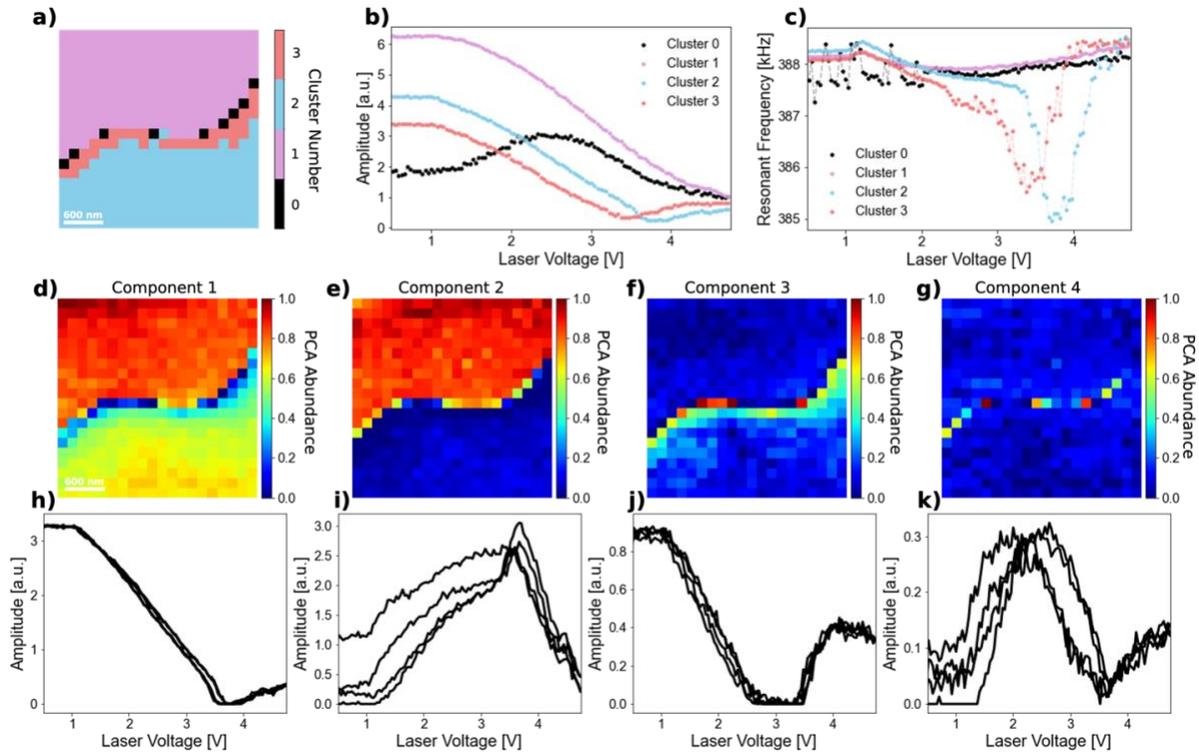

**Figure 3: K-means clustering and principal component analysis of thermal spectroscopy.** a) K-means clustering (k=4) of laser assisted thermal BE piezoresponse spectroscopy with corresponding b) amplitude and c) resonant frequency averages derived from clusters. Principal component analysis of laser assisted thermal BE piezoresponse spectroscopy with d-g) abundance maps and h-k) spectra for components 1-4, respectively.

Figure 2a-d, respectively. As such, Figure 2e shows offset tPFM amplitude versus laser voltage across the domain wall. In the upper domain (up polarization), a continuous decrease in amplitude is observed between a laser voltage of 1.5V and 4.75V. In contrast, the lower domain experiences a continuous decrease in amplitude to zero, then begins to increase with increasing temperature suggesting a ferroelectric to paraelectric transition. Similarly, the phase rotates by 180° (Figure 2f) in the lower domain congruent with the increased response. Importantly, with increasing distance from the domain wall within the lower domain, the onset of zero response appears at higher temperatures. These observations suggest the domain wall undergoes broadening, which is accompanied by domain dependent phase transition characteristics.

To gain further insight into spatial variability of the observed behaviors, we employ k-means clustering and principal component analysis (PCA) on the tPFM spectra. Briefly, k-means is an unsupervised machine learning algorithm that automatically assigns each pixel to a group

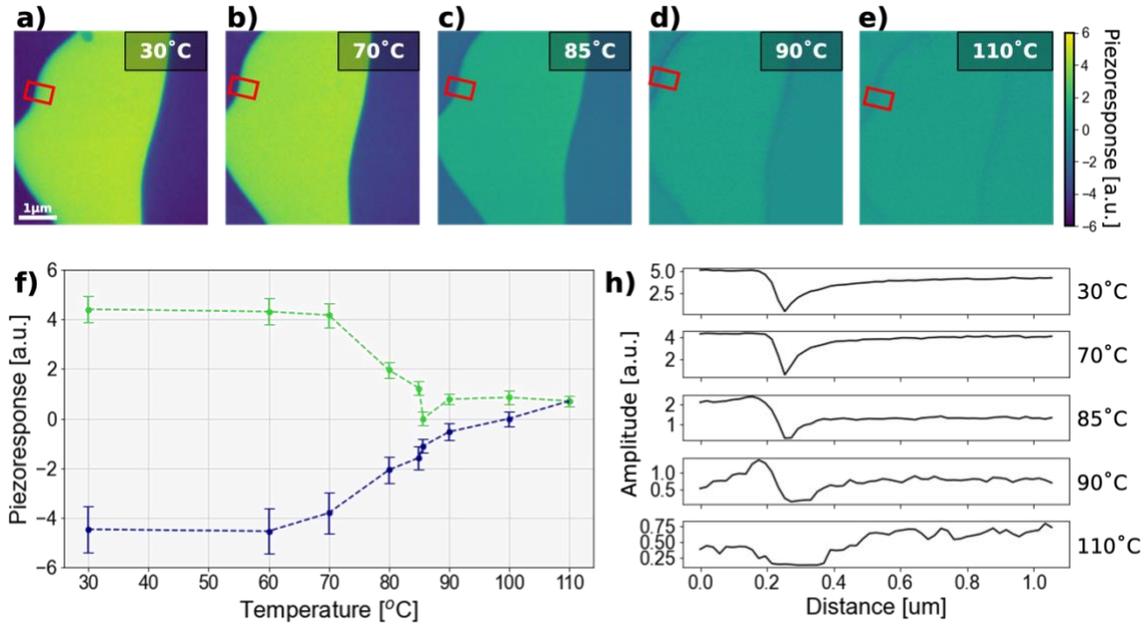

**Figure 4: Uniform heating band excitation piezoresponse force microscopy.** a-e) BE-PFM piezoresponse images of similar domain structure acquired at 30˚C, 70˚C, 85˚C, 90˚C, and 110˚C respectively. Note, piezoresponse drops by an order of magnitude at 110˚C. f) Histogram peaks and corresponding widths derived from green areas (up polarized domains, green line) and blue areas (down polarized domain, blue line) in panels (a-e) plotted versus temperature showing different transition temperatures in each domain. h) Average BE-PFM amplitude profiles across domain wall acquired from the red box in panels (a-e) versus temperature

(cluster) such that the within-cluster sum of squares is minimized. Comparatively, PCA is a linear decomposition technique used to decompose the dataset into the characteristic temperature dependent behaviors associated with position-dependent loadings. The application of these techniques on hyperspectral data is discussed in detail elsewhere.[42-46]

Here, using k-means clustering (k=4), four unique behaviors are observed where the spatial cluster assignments are shown in Figure 3a. The four clusters are located near the domain wall (black), at the domain wall (red), and in the up polarized (purple) and down polarized (blue) domains. The tPFM amplitude and resonant frequency averages pertaining to each cluster are plotted in Figure 3b,c respectively. As shown in Figure 3b, tPFM spectra acquired near the domain wall (cluster 0) have an increase in amplitude with a maximum at approximately 2.7V; tPFM spectra acquired at the domain wall (cluster 3) have an amplitude minimum at lower laser voltages as compared to the down polarized domain (cluster 2); and tPFM spectra acquired in the up polarized domain (cluster 1) have a continuously decreasing amplitude, similar to Figure 2e. Note, the amplitude response for all four clusters at the maximum temperature qualitatively approach the same value indicating at 4.75V the entire system has undergone a complex phase transition. Correspondingly, the decrease in resonance frequency (Figure 3c) observed near the domain wall at higher temperatures suggests a local softening of the material, consistent with previous reports of domain wall mechanics.[47]

Subsequently, we perform PCA on the tPFM amplitude with the component abundance maps and spectra shown in Figure 3d-g and Figure 3h-k, respectively. Component 1 captures the temperature induced response decay for both domains, while component 2 corresponds to the higher transition temperature located in the upper domain (up polarized), *i.e.*, combining component 1 and 2 qualitatively produces cluster 1 in Figure 3b. Component 3 corresponds to the domain wall broadening with increasing temperature, similar to Figure 2. Moreover, component 4 clearly captures the enhanced response observed near the domain wall. As such, the PCA results provide additional confirmation of the unique phase transition behavior derived from k-means, and together qualitatively show the nature of the complex phase transition present in SPS.

We further explore the phase transition behavior via uniform heating studies without the presence of the large local thermal field created in the UV laser assisted tPFM. Figure 4a-e shows band excitation PFM piezoresponse images from uniform heating studies at temperatures ranging from 30˚C to 110˚C. At lower temperatures, i.e. 30˚C and 70˚C (Figure 4a,b), the piezoresponse magnitude is relatively constant; however, at higher temperatures, i.e. 85˚C to 110˚C (Figure 4c-e), the response begins to rapidly decay, as expected for a ferroelectric-paraelectric phase transition. To better understand the phase transition behavior under uniform heating, we plot histograms (Supplementary Information) of the PFM piezoresponse images with the corresponding distribution maximum and FWHM (error bars) versus temperature shown in Figure 4f. Here, distinctly different transition behaviors between the up and down polarized domains are observed, similar to the tPFM, where the up polarized domains have a lower ferroelectric-paraelectric phase transition temperature. Lastly, we examine the domain wall behavior under uniform heating by extracting the domain wall width versus temperature in the region indicated by the red boxes in Figure 4a-e. As such, Figure 4g shows the average PFM amplitude across the domain wall. Note the continuous broadening of the wall with temperature, achieving ~150 nm at 110˚C.

Overall, tPFM and uniform heating PFM demonstrates intriguing phase transition behavior in SPS. Specifically, tPFM provides a continuous spectrum of thermally induced polarization dynamics where we observe a clear broadening of the domain wall and the phase transition in the down polarized (lower) domain propagates away from the domain wall. Correspondingly, the down polarized domain has a lower transition temperature as compared to the up polarized (upper) domain, which is further corroborated via uniform heating studies (Figure 4f). As such, these observations suggest domain dependent ferroelectric-paraelectric transition temperatures, *i.e.* the minimum piezoresponse as a function of temperature depends on the polarization orientation.

To gain further insight into the observed phenomena, we performed Finite Element Modeling reflecting the experimental conditions. Here, we aim to understand the temperature dependent polarization and effective piezoresponses in both a uniform and local heating (i.e. tPFM) geometry while considering the effects of rigid pinning on domain wall motion. For completeness, we also model the temperature dependent polarization, vertical displacement, and effective piezoresponse at the domain wall versus a uniformly polarized surface within a local heating PFM geometry. Accordingly, we establish a decoupled model for thermal PFM. We assume that the temperature distribution inside the material is established due contact with the heated AFM probe and is calculated self-consistently. The equilibrium temperature distribution $T(\vec{r})$ satisfies a Laplace equation inside the SPS region:

$$\Delta T(\vec{r}) = 0. \tag{1}$$

Due to the very high heat conductivity of the metallic tip in comparison with a moderate conductivity of SPS and very low conductivity of the air, one can neglect the heat flux between the SPS and air and assume fixed temperature $T_d$ at the tip-surface contact area. Far from the heated tip the temperature tends to $T_0$. Using a disk-plane model of the SPM tip,[48, 49] we substitute the tip apex by a uniformly heated disk of radius $R \sim (5 - 50)$ nm, which is in a perfect thermal contact with the SPS surface $z = 0$.

The polarization and strain fields are calculated numerically as a non-local problem assuming the known temperature distribution as a solution of coupled Ginzburg-Landau and elastic equations. The relevant details including constitutive relations and materials parameters are provided in the Supplementary Information. Briefly, the equilibrium spatial distributions of the electric field $E_i$ and out-of-plane ferroelectric polarization component $P_3$ obey the coupled problem consisting of Poisson equation for electric potential $\varphi$ and Landau-Ginzburg-Devonshire (LGD) equation for the $P_3$:

$$\Delta\varphi = \frac{1}{\varepsilon_0 \varepsilon_b}\frac{\partial P_3}{\partial z}, \qquad (2a)$$

$$[\alpha(T) - Q_{ij}\sigma_{ij}]P_3 + \beta P_3^3 + \gamma P_3^5 - g_{11}\frac{\partial^2 P_3}{\partial z^2} - g_{44}\left(\frac{\partial^2 P_3}{\partial x^2} + \frac{\partial^2 P_3}{\partial y^2}\right) = \mu\frac{dT}{dz} + E_3. \qquad (2b)$$

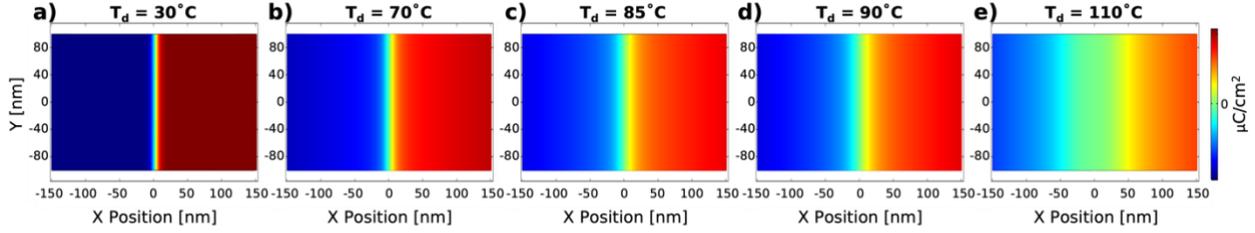

**Figure 5:** The top view of polarization $P_3$ across 180-degree domain wall (DW) at the SPS surface calculated for different tip temperatures 30, 70, 85, 90, and 110°C. Far from the DW boundary one can see "up" (red) and "down" (blue) domains. Ambient temperature $T_0 = 20°C$, $V = 0$. Note, tip-surface contact radius to the film thickness ratio $R/h = 0.09$, and $\mu = -6.0 \times 10^{-5}$ V/K were determined from the best comparison to experimental data.

Here, SPS layer is located at $0 \leq z \leq h$, $\varepsilon_b$ is a background permittivity,[50] $[\alpha_T(T(\vec{r}) - T_C)$, where $T_C \approx 67°C$ is the Curie temperature, $Q_{ij}$ are electrostriction tensor components, $\sigma_{ij}$ are elastic stresses, and $\mu$ is the coefficient of thermo-polarization effect.[51] The coefficient $\mu$ is the diagonal component of the second rank tensor $\mu_{ij}$, which values can be estimated as proportional to the product of the flexoelectric $F_{ij}$ and linear thermal expansion $\beta_{ij}$ coefficients, $\mu_{ij} \cong F_{ijkl}\beta_{kl}$.[i] Assuming the dominant role of the thermo-polarization effect over the flexoelectric effect, the latter is omitted in this work. The electric potential is $V$ at the tip-surface contact area and vanishes at bottom electrode, The potential $\varphi$ obeys the Laplace equation outside the ferroelectric. As such, the equations of state for elastic stresses $\sigma_{ij}$ and strains $u_{ij}$ are:

$$u_{ij} = s_{ijkl}\sigma_{kl} + \beta_{ij}(T - T_0) + Q_{ijkl}P_k P_l, \qquad (3)$$

Analysis of the SPS polarization and effective piezoresponse for different tip voltages revealed that starting from $|V| > 10$ mV we are already in the "strong modulation" regime, i.e., such small voltages lead to domain wall motion and nanodomain nucleation under the tip (see

---

[i] The estimate follows from relations $\delta u_{ij} \cong \delta T \beta_{kl}$ and $\delta P_i \cong F_{ijkl}\frac{\partial \delta u_{kl}}{\partial x_j} \cong F_{ijkl}\beta_{kl}\frac{\partial \delta T}{\partial x_j}$.

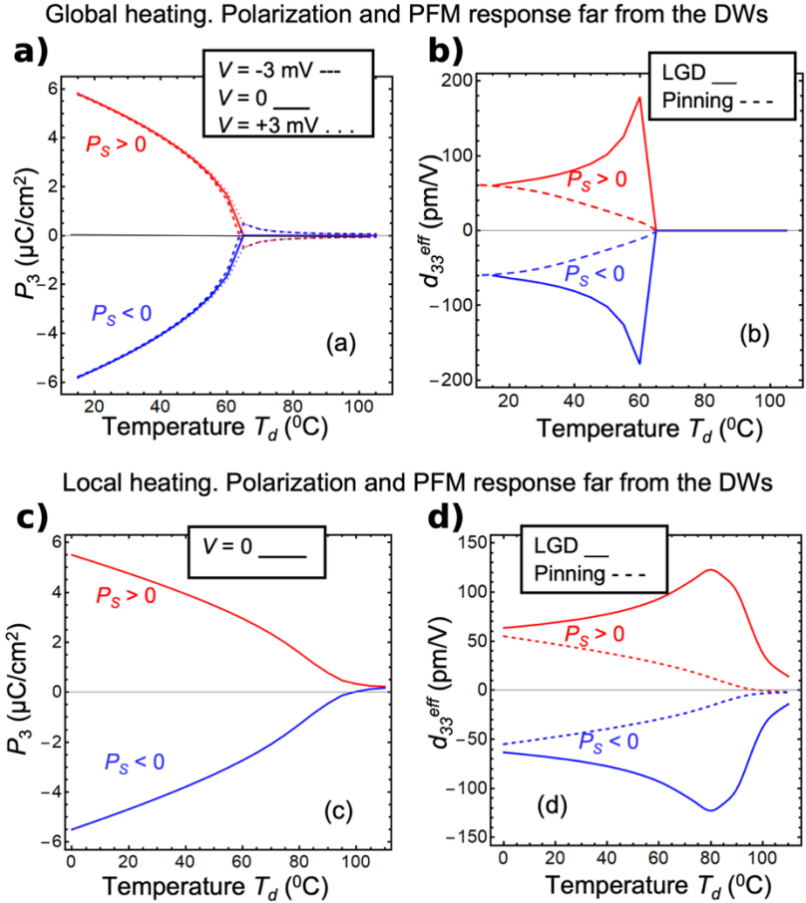

**Figure 6:** Surface polarization $P_3$ (a, c) and effective piezoresponse $d_{33}^{eff}$ (b, d) under the tip apex calculated versus the tip temperature $T_d$ for uniform heating (a, b) and local heating (c, d). The tip is located very far from the domain walls. Different directions of the spontaneous polarization $P_S$ correspond to "up" (red curves) and "down" (blue curves) domains. For plots (c, d) the ambient temperature $T_0 = 20^0\text{C}$. The tip voltage $V$ is either zero or very small (see legends at the plots). Solid and dashed curves in the plots (b) and (d) are calculated in the LGD model without pinning and allowing for the "rigid" pinning, respectively.

Figure S1). This happens because the SPS Curie temperature $T_C$ is close enough to room temperature where local heating at e.g., 30°C transforms the region under the tip very close to the paraelectric phase. Furthermore, small voltages cause strong polarization and piezoresponse modulation near the paraelectric phase boundary.

The top view of polarization $P_3$ across a 180-degree domain wall (DW) at the SPS surface was calculated for different tip temperatures 30, 70, 85, 90, and 110°C (Figure 5a-e, respectively) with ambient temperature $T_0 = 20°C$. Note, far from the DW boundary one can see "up" (red) and "down" (blue) domains (see Supplementary Information for further details). Here, the heated and biased tip is placed on the domain wall resulting in clear temperature-induced domain wall broadening (Figure 5, compare the DW width from the left to right) similar to the experimental observations in Figure 4h.

To provide numerical insight into the materials responses, surface polarizations $P_3$ and effective piezoresponses $d_{33}^{eff}$ under the tip apex were calculated versus the tip temperature $T_d$ for uniform (Figure 6a-b) and local heating (Figure 6c-d) using the LGD model. Moreover, the different directions of spontaneous polarization $P_S$ correspond to "up" (red curves) and "down" (blue curves) domains, where the tip voltage $V$ is either zero or very small (±3mV). Here, the tip is located far from the domain wall. The effective piezoresponse under uniform heating is shown in Figure 6b with and without pinning (dashed and solid curves, respectively). Without pinning, the effective piezoresponses are non-monotonic functions of temperature and possess a sharp maximum at approximately 60°C (slightly less than the bulk Curie temperature), which is associated with the increase of dielectric susceptibility near the apparent paraelectric transition at $T_C$. Importantly, without pinning, the domain walls move at infinitely small tip voltages with non-monotonic growth near Tc originating from the maxima of dielectric permittivity. In contrast, the dashed curves in Figure 6b are calculated allowing for "rigid" pinning assuming the domain wall cannot move at voltages smaller than the threshold. In this approximation, the signal monotonically decreases with temperature resembling the polarization behavior shown in Figure 6a. Specifically, the PFM response in this case is proportional to the electrostriction terms in Eq.(3) linearized with respect to electric field, that gives the combination of piezoelectric tensor components, $d_{ijm} = 2Q_{ijkl}P_k^S \varepsilon_{lm}$. The latter are in turn proportional to the spontaneous polarization $P_k^S$ and dielectric permittivity $\varepsilon_{lm}$. Since the rigid pinning prevents the local increase of the permittivity near the wall, PFM response curves are proportional to the spontaneous polarization shown in Figure 6a.

In comparison, under local heating, the spontaneous polarization disappears under the tip for the tip temperatures $T_d > T_{cr}$, due to the local transition to a paraelectric phase (see Figure 6c). Importantly, the temperature $T_{cr}$ is approximately 110ºC, however some thermo-induced polarization remains. Here, the difference between "up" and "down" domains originates from the thermo-polarization effect, mathematically expressed by the term $\mu \frac{dT}{dz}$ in LGD equations for polarization (see **Appendix A** for details).

The effective piezoresponse under local heating with and without pinning are shown in Figure 6d. In the unpinned case (solid lines), similar to uniform heating, the effective piezoresponses are non-monotonic functions of temperature and reveal a maximum at 90°C associated with the increase in dielectric susceptibility near the apparent paraelectric transition. However, in the rigidly pinned case (dashed lines), the effective piezoresponses monotonically decrease with temperature resembling the polarization behavior shown in Figure 6c. Note, the local piezoresponse curves were calculated from the displacement profiles for very small voltages $V \sim \pm 1$ mV, using the expression of a symmetrized derivative $d_{33}^{eff} \approx \frac{U_3(V) - U_3(-V)}{2V}$. The small voltages and symmetrization minimize the effect of possible domain nucleation in applied E-field. In essence, the modeled enhanced response in both heating geometries in the unpinned case qualitatively match the observed response near the domain walls (Figure 3), suggesting the observed enhanced response is due to a lower degree of pinning near the domain walls.

Accordingly, next we investigate the thermally induced responses near the domain wall. As such, the surface polarization, displacement profiles and effective piezoresponse across the straight 180-degree domain wall (DW) in thick SPS are shown in Figure 7a-c, respectively. Here, local heating leads to temperature-induced domain wall broadening (compare blue, green and red curves in Figure 7a) and further shape changes (see teal and violet curves in Figure 7a), similar to Figure 5 and the observed response if Figure 4h. It is seen that the domain wall broadens monotonically for $T_d < T_{cr} \sim 90°C$, and at $T_d > \Delta T_{cr}$ where the paraelectric region appears under the tip (compare the violet curve with the others). Corresponding profiles of the surface displacement are shown in Figure 7b, where the principal changes in the profile shape happens at lower temperatures, at $T_d > 50°C$ (compare the teal and violet curves with the others). The local effective piezoresponse curves were calculated from the displacement profiles for very small voltages $V \sim \pm 1$ mV, using the expression of a symmetrized derivative $d_{33}^{eff}(x, V) \approx \frac{U_3(x,V) - U_3(x,-V)}{2V}$ to minimize the effect of possible domain nucleation and/or DW motion in applied E-field, similar to before. The DW motion starts in very small E-fields, since we use LGD-based FEM without pinning effects. The local piezoresponse curves are shown in Figure 7c. It is seen from the plot that local heating drastically changes the piezoresponse profiles, since pronounced maximum and minimum appear near the different sides of the DW. The entire profile is antisymmetric with respect to the DW plane, and the paraelectric "hole burning" (violet curves) exists for both straight and curved walls.

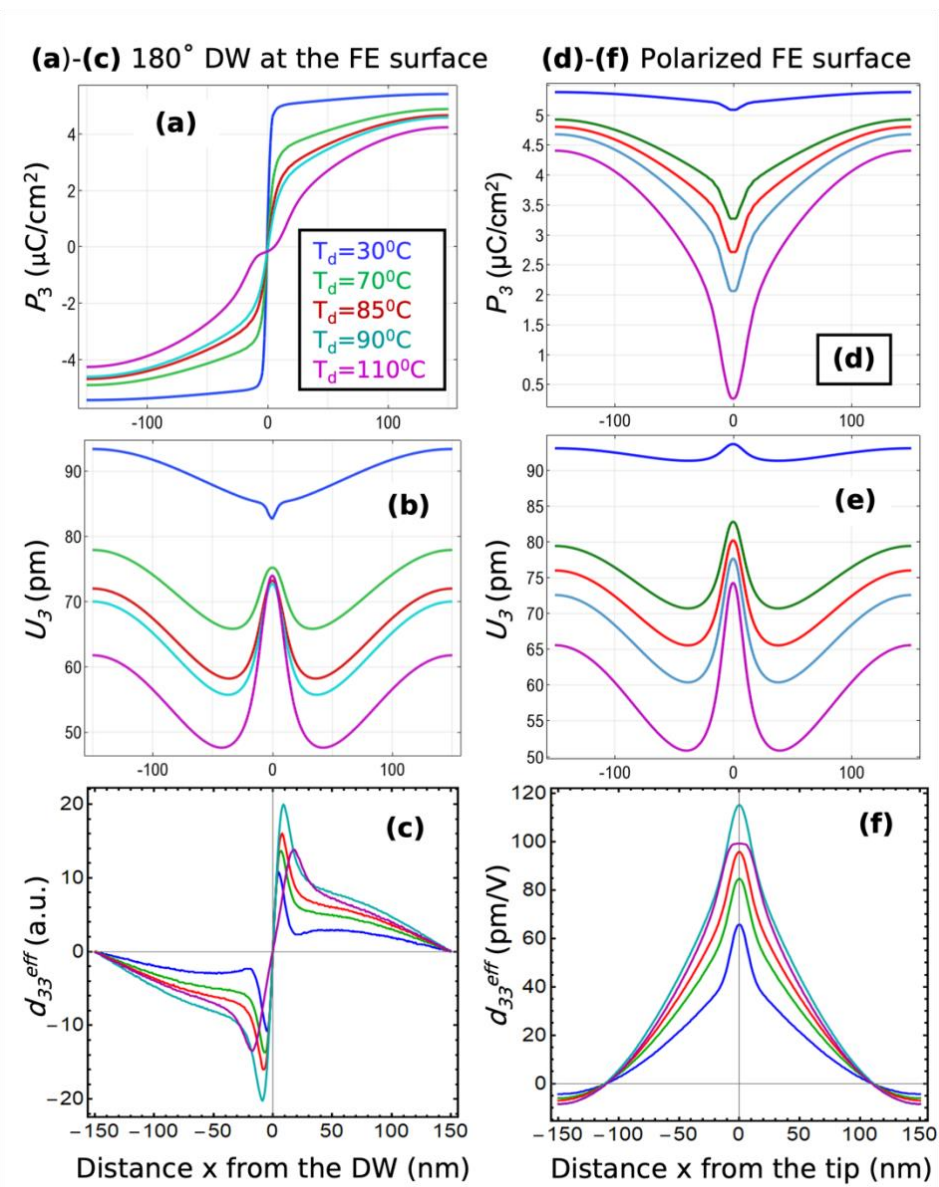

**Figure 7:** Polarization $P_3$ (a, d), vertical displacement $U_3$ (b, e) and effective piezoresponse $d_{33}^{eff}$ (c, f) profiles of the SPS surface across the straight 180-degree domain walls (a-c) and for the homogeneously polarized SPS surface (d-f) calculated for different tip temperatures 30 (blue curves), 70 (green curves), 85 (red curves), 90 (teal curves), and 110 ºC (violet curves). Ambient temperature $T_0 = 20^0C$. The tip voltage $V$ is either zero or very small (see legends at the plots).

Lastly for comparison, we identify the surface polarization, displacement profiles and effective piezoresponse of thick homogeneously polarized SPS as shown in Figure 7d-f, respectively. Local heating leads to the temperature-induced decrease of the spontaneous polarization (compare different curves in Figure 7d), but its disappearance is expected at temperatures higher than 110ºC (see violet curves in Figure 7d). Corresponding profiles of the surface displacement, shown in Figure 7e, reveal that the principal changes in the profile shape happens much earlier, at $\Delta T > 70$ K (compare the teal and violet curves with the others). The local piezoresponse curves, shown in Figure 7f, illustrate that the local heating drastically changes the piezo-response profiles where a pronounced maximum appears under the heated tip. Note, the entire profile is symmetric with respect to the tip axis.

To summarize, here we report the experimental realization of the temperature-assisted piezoresponse force microscopy. In this method, the local electromechanical response is measured as a function of tip temperature, providing insight into the local temperature-induced polarization dynamics and phase transitions associated with changes in the polar behavior. For SPS, we discover temperature-induced broadening and effective transition temperature lowering at the domain wall. Moreover, we observe distinctively different phase transition behavior between up and down polarized domains, suggesting domain dependent Curie temperatures. Intriguingly, both tPFM and uniform heating studies suggest the presence of weak (less then 10% of RT value) constant polarization above the (nominal) bulk phase transition associated with the preexisting domains where the polarization is zero at the preexistent domain walls. While we tentatively identify this behavior with surface chemical pinning, it is clearly a subject for further investigations.

Overall, the tPFM can be broadly used for the analysis of thermal phase transitions in materials exhibiting local electromechanical responses. This includes classical ferroelectric materials, but also ionic materials such as those used in energy and information technologies, in which electromechanical response is associated with electrochemical phenomena in the tip-surface junction.[52-55] Similarly, this method can be extended to probing electrostatic responses associated with surface charging, directly related to the charge injection and surface electrochemistry of materials. As such, it provides a new and powerful venue for the exploration of phase transitions, and bulk and surface electrochemistry on the nanoscale.

# METHODS

*Laser Assisted Thermal Band Excitation Piezoresponse Force Microscopy*

Laser assisted tPFM measurements were deployed via an Oxford Instruments Asylum Research Cypher atomic force microscope equipped with a "BlueTherm", 40 mW laser diode. All PFM experiments used Budget Sensor Multi75E-G Cr/Pt coated AFM probes (~3 N/m). Band excitation PFM was collected using an AC voltage of 1V, while thermal spectroscopy was collected using laser voltages ranging from 0V-4.75V at 1Hz cycling frequency. All measurements were performed in ambient conditions in standard humid conditions (~20-50%).

# MODELING

*Finite Element Modeling*

Finite Element Modeling was done in COMSOL@MultiPhysics software, using electrostatics, solid mechanics, and general math (PDE toolbox) modules. A portion of results were analyzed and visualized in Mathematica 12.2 (https://www.wolfram.com/mathematica)

# ACKNOWLEDGEMENTS

This effort (K.K., S.V.K.), including the data analysis and interpretation, was supported as part of the center for 3D Ferroelectric Microelectronics (3DFeM), an Energy Frontier Research Center funded by the U.S. Department of Energy (DOE), Office of Science, Basic Energy Sciences under Award Number DE-SC0021118. The PFM experiments were performed and partially supported (S.J., P.M.) at Oak Ridge National Laboratory's Center for Nanophase Materials Sciences (CNMS), a U.S. Department of Energy, Office of Science User Facility. Theory (A.N.M.) was supported by the National Academy of Sciences of Ukraine (the Target Program of Basic Research of the National Academy of Sciences of Ukraine "Prospective basic research and innovative development of nanomaterials and nanotechnologies for 2020 - 2024", Project № 1/20-H, state registration number: 0120U102306) and received funding from the European Union's Horizon 2020 research and innovation programme under the Marie Skłodowska-Curie grant agreement No 778070.

# DATA AVAILABILITY

The data that support the findings of this study are available from the corresponding author upon reasonable request.